%% file: pqas_draft.tex
\documentclass[aps,pra,twocolumn,preprintnumbers,amsmath,amssymb,superscriptaddress]{revtex4-2}
\usepackage{graphicx}
\usepackage[caption=false]{subfig}
\usepackage{epsfig}
\usepackage{dcolumn}
\usepackage{bm}
\usepackage{color}
\usepackage{float}
\usepackage[colorlinks]{hyperref}
\usepackage{verbatim}
\usepackage{algorithm}
\usepackage{algorithmic}

\hypersetup{citecolor = blue}

\def\be{\begin{equation}}       \def\ee{\end{equation}}
\def\bea{\begin{eqnarray}}      \def\eea{\end{eqnarray}}
\def\ba{\begin{array}}
\def\ea{\end{array}}
\def\bnum{\begin{enumerate} }
\def\enum{\end{enumerate}}

\def\=>{\Rightarrow}
\def\>{\rightarrow}

\def\eye2{Fathbb{I}}

\renewcommand{\v}[1]{{\bf #1}}

\renewcommand{\>}{\rangle}

\usepackage{listings}
\usepackage{color}
\definecolor{lightgray}{gray}{1}

\begin{document}

\title{Neural Predictor based Quantum Architecture Search}

\author{Shi-Xin Zhang}
\affiliation{Institute for Advanced Study, Tsinghua University, Beijing 100084, China}
\affiliation{Tencent Quantum Laboratory, Tencent, Shenzhen, Guangdong 518057, China}
\author{Chang-Yu Hsieh}
\email{kimhsieh@tencent.com}
\affiliation{Tencent Quantum Laboratory, Tencent, Shenzhen, Guangdong 518057, China}
\author{Shengyu Zhang}
\affiliation{Tencent Quantum Laboratory, Tencent, Shenzhen, Guangdong 518057, China}
\author{Hong Yao}
\email{yaohong@tsinghua.edu.cn}
\affiliation{Institute for Advanced Study, Tsinghua University, Beijing 100084, China}

\begin{abstract}
	
Variational quantum algorithms (VQAs) are widely speculated to deliver quantum advantages for practical problems 
under the quantum-classical hybrid computational paradigm in the near term. Both theoretical and practical developments of VQAs share many similarities with those of deep learning. For instance, a key component of VQAs is the design of task-dependent parameterized quantum circuits (PQCs) as in the case of designing a good neural architecture in deep learning. Partly inspired by the recent success of AutoML and neural architecture search (NAS), quantum architecture search (QAS) is a collection of methods devised to engineer an optimal task-specific PQC. It has been proven that QAS-designed VQAs can outperform expert-crafted VQAs under various scenarios. In this work, we propose to use a neural network based predictor as the evaluation policy for QAS. We demonstrate a neural predictor guided QAS can discover powerful PQCs, yielding state-of-the-art results for various examples from quantum simulation and quantum machine learning. Notably, neural predictor guided QAS provides a better solution than that by the random-search baseline while using an order of magnitude less of circuit evaluations.
Moreover, the predictor for QAS as well as the optimal ansatz found by QAS can both be transferred and generalized to address similar problems. 
	
\end{abstract}

\date{\today}
\maketitle

\section {Introduction}
Variational quantum algorithms \cite{Cerezo2020b,Bharti2021} (with a relatively low consumption on quantum resources) are readily accessible in the noisy intermediate scale quantum (NISQ) era \cite{Preskill2018} and show some promising signs for practical quantum advantage in the near future.
Some prominent examples of VQA include variational quantum eigensolvers (VQE) \cite{Peruzzo2014, OMalley2016, McClean2016,Liu2019b,  McArdle2020, Grimsley2019}, quantum approximation optimization algorithms (QAOA) \cite{Farhi2014,Hadfield2017,Zhou2020, Arute2020, McClean2020} and variational quantum machine learning (QML) \cite{Biamonte2017a, Farhi2018,Mitarai2018, Liu2018d, Lloyd2018, Verdon2019, Havlicek2019, Schuld2019, Benedetti2019a,Benedetti2019b,Cong2019,Verdon2019,Schuld2020, Huang2020a}.
Such parameterized quantum circuit based variational algorithms have been exploited to solve many tasks that are considered hard for classical computers, such as integer factoring, combinatorial optimization, quantum Hamiltonian ground state solver, and quantum dynamics simulator.

The conventional VQA requires a fixed quantum structure ansatz where the trainable parameters are iteratively adjusted via a classical optimizer in order to minimize an objective function.  Some famous PQCs include QAOA ansatz for combinatorial problems \cite{Farhi2014}, hardware efficient ansatz \cite{Kandala2017} and unitary coupled clusters (UCC) ansatz \cite{Peruzzo2014, Wecker2015a, OMalley2016} for VQE. There are also extensive studies on the expressive power and trainability on these ansatzes \cite{Sim2019, Farhi2020, Akshay2020,Rasmussen2020, Akshay2020b}. However, it still remains open on how to discover specifically tailored quantum ansatzes for different tasks. Partly inspired by neural architecture search (NAS) from the AutoML community \cite{Yao2018, Elsken2019a, Wistuba2019, Ren2020}, we introduced the concept of quantum architecture search (QAS), which refers to a collection of effective methods that systematically search for an optimal quantum circuit ansatz for a given problem in Ref. \cite{Zhang2020b}. Due to the apparent analogy between variational quantum circuits and neural networks, approaches developed in NAS have been adapted to QAS in the quantum computing domain.
Representative works have utilized ideas including evolutionary/genetic algorithms \cite{LasHeras2016, Li2017c,Cincio2018a, Rattew2019,Cincio2020,Chivilikhin2020, Lu2020}, reinforcement learning (RL) approaches  \cite{Fosel2018a, Niu2019}, greedy algorithms \cite{Ostaszewski2019, Grimsley2019,Tang2019,Li2020}, and differentiable architecture search \cite{Zhang2020b}.

While QAS is an elegant idea, it faces an important challenge of exploring and evaluating many quantum ansatzes during the training. Yet this computational bottleneck is intrinsic to all design-by-search methodologies including NAS in deep learning. In order to enhance search efficiency for NAS,  there are two mainstream evaluation policies widely adopted in NAS. The first one is weights sharing, where trainable parameters are reused instead of standalone training for each ansatz. Such weights sharing policy is utilized in one-shot search \cite{Bender2018, Luo2019,Guo2019} and DARTS \cite{Liu2019a} in NAS, and the same idea has been exploited in the corresponding QAS frameworks, quantum circuit architecture search \cite{Du2020a} and differentiable quantum architecture search \cite{Zhang2020b}, respectively. The second type of evaluation policy is to evaluate the fitness of an architecture by a meta machine learning model. Such predictor-based methods constitute another well-established and actively researched subfield in NAS \cite{Deng2017b, Liu2018c, Dai2019, Wang2018,  White2019, Shi2019, Wen2020,Ning2020,Chen2020d, White2020a,Lukasik2020,Dudziak2020,Mauch2020}; yet, to the best of our knowledge, a predictor-based QAS framework has not been explored so far. In this work, we introduce the first predictor-based QAS.
We train a neural predictor to directly gauge the performance of candidates of quantum circuits using only the structure of quantum ansatz. This predictor is then integrated into a QAS workflow (to be elucidated) and can substantially accelerate the search process.

The main contributions of the present work are summarized below.
\begin{enumerate}
	\item We introduce meta network/neural predictor to a QAS framework, and achieve a dramatic boost on search efficiency compared to a random search baseline.

	\item We demonstrate that the proposed neural predictor based QAS performs competitively in some VQE and QML tasks as it efficiently discovers state-of-the-art quantum architectures.

	\item We also show the optimal circuit structure suggested by QAS can be transferred to problems of different size. We develop a sophisticated transfer protocol incorporating beam search and evolutionary techniques to achieve this.
\end{enumerate}

\section{Related Work}

\subsection{Predictor based NAS}

NAS \cite{Yao2018, Elsken2019a, Wistuba2019, Ren2020} is a recently emerging and rapidly developing field.
It aims at, automatically, searching for optimal neural networks for some given tasks without relying on handcrafted design and expert guidance. An effective NAS workflow is composed of several ingredients with the sampling strategy and evaluation protocol being among the most important. In short, sampling strategy refers to a customized recommendation of candidate neural networks, whose fitness for a specific task should be evaluated (eg. training and checking accuracy on some validation/test sets). Since the search space is exponentially large, we need an efficient method to traverse and sample candidate architectures. The sampling strategies that have been previously attempted include random search \cite{Li2019c}, local/greedy search \cite{Huang2018a,White2020,Li2019}, evolutionary/genetic type search \cite{Real2017a, Xie2017, Liu2018a, Real2019a, Stanley2019}, RL based search \cite{Zoph2017, Baker2017, Cai2018, Zoph2018}, Bayesian optimization search \cite{Dai2019, White2019, Shi2019} and so on. While sampling is essential for NAS, the computational bottleneck is really the evaluation part as it is time consuming  to train individual networks (from scratch) on a large dataset. In fact, early NAS works \cite{Zoph2017} using such plain evaluation methods with RL or evolutionary engine often take thousands of GPU hours before identifying an optimal neural network architecture.

There are two approaches to lower the cost of individual evaluations. The first one is weights sharing. In this setup, all candidate networks are organized within a super network, which can be trained either by training the entire network or by training sampled subnetworks in each epochs. After training the super network, we can evaluate each candidate subnetwork by regarding it as the child of the super network and the weights of such candidates are inherited from the super network directly without further tuning. Therefore, the evaluation on the candidate network is as simple as a forward pass for inference.  Such parameter sharing setup is inspired from one-shot NAS \cite{Bender2018, Luo2019,Guo2019} and has also become popular for DARTS \cite{Liu2019a, Casale2019, Cai2019, Xie2019}.
Parameter sharing is not perfect, though. As there is actually no theoretical guarantee on the accuracy correlation between subnetworks with inherited weights and optimal weights from individual training.

This work focuses on the second strategy to improve the evaluation methodology. Instead of training a candidate network and evaluating its performance on some validation dataset, we directly build machine learning (ML) models to predict the network performance based on the network structure alone, i.e. without specifying trainable weights. If the prediction accuracy manifests non-trivial correlation with the ground truth, then such predictor based evaluation method may greatly ameliorate NAS efficiency. We denoted works along this line as predictor based NAS, in which the so-called ``predictor" is a regression model \cite{Deng2017b, Liu2018c, Dai2019, Wang2018,  White2019, Shi2019, Wen2020,Ning2020,Chen2020d, White2020a,Lukasik2020,Dudziak2020,Mauch2020}. Some works take a further step by constructing variational autoencoder (VAE) \cite{Kingma2013} for neural architectures and trains a predictor with input from the latent space of VAE \cite{Luo2018, Zhang2019c,Friede2019,  Cheng2020,  Li2020c, Zheng2020,Tang2020}. Such predictor can also be transferred to infer other metrics of the network such as latency or FLOPs \cite{Dai2019, Dudziak2020}. It is worth noting that predictor based NAS can also be combined with weight sharing tricks, where the predictor is actually trained with fitness label obtained from one-shot setup \cite{Luo2018,Zhang2019c,Shi2019,Chen2020d}.

\subsection{Variational Quantum Algorithms}

The paradigm of variational quantum algorithms (VQA) \cite{Cerezo2020b, Bharti2021} is widely speculated to deliver practical quantum advantages with NISQ devices \cite{Preskill2018} in the near term. The hybrid quantum-classical implementation of VQA bears many similarities to deep learning in classical computations. Various VQA have been designed for problems in a vast array of fields.
Some typical examples include determining the ground and low excited states of a quantum Hamiltonian \cite{Cao2019,McArdle2020,  Peruzzo2014,McClean2016, Liu2019b, Higgott2019,Nakanishi2018,Hsieh2019}, simulating quantum dynamics \cite{Li2017b, Yuan2019, McArdle2019,Cirstoiu2020,  Lin2020, Lau2021, Endo2020a, Benedetti2020,Bharti2020a,Barison2021}, preparing Gibbs thermal states \cite{Liu2019, Wang2020b, Chowdhury2020}, supervised and unsupervised machine learning \cite{Biamonte2017a, Farhi2018,Mitarai2018, Liu2018d, Lloyd2018, Verdon2019, Havlicek2019, Schuld2019, Benedetti2019a,Benedetti2019b,Cong2019,Verdon2019,Schuld2020, Huang2020a}, compiling quantum circuits \cite{Khatri2019, Heya2018, Sharma2020}, solving classical combinatorial optimizations \cite{Farhi2014,Hadfield2017,Zhou2020, Arute2020, McClean2020}, factoring integers \cite{Anschuetz2019, Saxena2020, Karamlou2020}, and solving linear or differential equations \cite{Huang2019,Xu2019c,Bravo-Prieto2019,Kyriienko2020a,Lubasch2020} etc.  VQA requires a fixed PQC ansatz at the beginning stage. Such fixed ansatz are usually designed to accommodate limitations of current hardware \cite{Kandala2017} or are inspired from well-established theoretical techniques, such as quantum annealing, unitary coupled cluster etc. \cite{Yung2014, Peruzzo2014, Wecker2015a, OMalley2016, Farhi2014}.

Despite wide application range and initial successes, pushing for quantum advantages with VQA faces many challenges such as noise-induced decoherence, barren plateaus \cite{McClean2018, Wang2020} that derail the training of parameters, and reachability deficits with certain fixed ansatz \cite{Sim2019, Akshay2020,Akshay2020b,  Farhi2020}. Although there are various proposals \cite{Li2017b, Temme2017,Grant2019, Patti2020, Cerezo2020a} on alleviating these issues, it is extremely difficult to fully resolve them as long as the circuit ansatz is fixed at the beginning. 

\subsection{Quantum Architecture Search}

As discussed in the last section, a parametrized quantum circuit is required as an \textit{ansatz} for VQA. A badly designed ansatz could possess limited expressive power and/or entangling capacity, leaving the global minimum for an optimization problem out of reach. Furthermore, such ansatz may be more susceptible to noises \cite{Franca2020},  wastes quantum resource or leads to barren plateau that frustrates the optimization procedure \cite{McClean2018, Wang2020}.

Therefore, a systematic approach to search for optimal circuit ansatz is desired, and we denote such workflow as ``quantum architecture search" \cite{Zhang2020b}. The aim of the QAS is to recommend tailored quantum circuits for a given problems such that it not only  minimizes a loss function, but also satisfies a few other constraints imposed by the hardware connectivity among qubits, native quantum gate set, quantum noise model, training loss landscape and other practical issues.

Previous QAS works have heavily borrowed ideas from NAS. More specifically,  greedy methods \cite{Ostaszewski2019, Grimsley2019,Tang2019,Li2020}, evolutionary or genetic methodologies \cite{LasHeras2016, Li2017c,Cincio2018a, Rattew2019,Cincio2020,Chivilikhin2020, Lu2020}, RL engine based approaches \cite{Fosel2018a, Niu2019}, Bayesian optimization \cite{Pirhooshyaran2020}, one-shot search \cite{Du2020a} and gradient based methods \cite{Zhang2020b} have all been adopted to discover better circuit ansatz for VQA. However, as far as we know, predictor based evaluation strategy has not been applied toward quantum circuit design. Since predictor based NAS has been empirically demonstrated to be highly efficient, one may anticipate predictor based QAS to hold similar performance boost for VQA in the NISQ era.

\section{Methods}

In this section, we describe the essential technical ingredients of our proposed neural predictor based QAS workflow. To facilitate the discussion, we layout a few definitions. Below, we use $N$ to denote the number of qubits in a circuit, $t$ to denote the number of types of quantum gate primitive in the search space, and $n_t$ to denote the total number of quantum primitives in one circuit.

\subsection{Search space for QAS}

We adopt two distinct representations (list of gates and image of a circuit) to denote each candidate circuit. In particular, the list representation comes with a strict syntax delineating a quantum circuit in terms of a sequence of applied quantum gates. Each list is a set of tuples, and each tuple encodes a quantum gate (referenced to a given gate set) and positional information (i.e. qubits numbered in a certain way).  For instance, tuple $(3,1,2)$ indicates a 3rd type of two-qubit quantum gates acting on the first and second qubit in the circuit. To reconstruct a corresponding quantum circuit from a list, the gates should be sequentially placed onto an initially empty circuit, according to the given tuple sequence. Note a circuit ansatz may have multiple list representations. One possibility is that the order of gate placements can be freely exchanged as long as gates encoded by the two tuples commute.  This syntax, inherently, gives a legitimate search space for circuit ansatz. Hence, in this work, sampled circuits are generated in this list representation.

As for the quantum gate sets, we choose different primitives for different problems. Some common examples include non-parameterized gate such as Hadamard gate $H$; single qubit gates with trainable parameters such as rotation gate $R_x=e^{-i\theta X/2}$ and, similarly, for $R_y, R_z$, where $X,Y,Z$ are corresponding Pauli matrices; and parameterized two-qubit gate such as $XX=e^{-i\theta X_1X_2/2}$ with counterparts for $YY,ZZ$ gate as well as parameterized SWAP gate $\text{SWAP}^{\theta}=e^{-i\theta \text{SWAP}/2}$ where $\text{SWAP}_{12}=I_1I_2+X_1X_2+Y_1Y_2+Z_1Z_2$.

Now, let us elaborate on the sampling strategy considered in this work.  First, we comment on the naive method of random sampling. After fixing the number of qubits and the total number of quantum gates in a circuit, one then can randomly sample list representation of quantum circuit ansatz.  This simple strategy is problematic and inferior in several ways. For instance, gate layout for randomly sampled circuits are highly ``chaotic" and usually incur severe issues of barren plateau since the circuits somehow behave like random unitaries drawn from the Haar measure. Besides, such randomly generated circuit ansatz is often deep as the arrangement of quantum gates is sparse. Lastly, random ansatz is not amenable to further utilization in the sense of generalization and transferability to problems of different size since there is no obvious pattern for extraction. Therefore, we devise two pipelines for circuit sampling that substantially alleviate aforementioned concerns.

The two sampling pipelines are gatewise generation and layerwise generation, respectively. In the first pipeline, we construct the circuit gate by gate by specifying their positions and types, while in the second pipeline, we construct the circuit by iteratively adding half-layers. Namely, whenever we pick a type of quantum gate we have to apply it on the set of even qubits or odd qubits. Both pipelines further incorporate additional techniques such as hierarchical generation and gate correlation enforcement. See Supplemental Materials for the details of these two sampling pipelines and the consideration on design of circuit search space.

\subsection{Representation of the quantum circuit}

\begin{figure}[t]
	\includegraphics[width=0.46\textwidth]{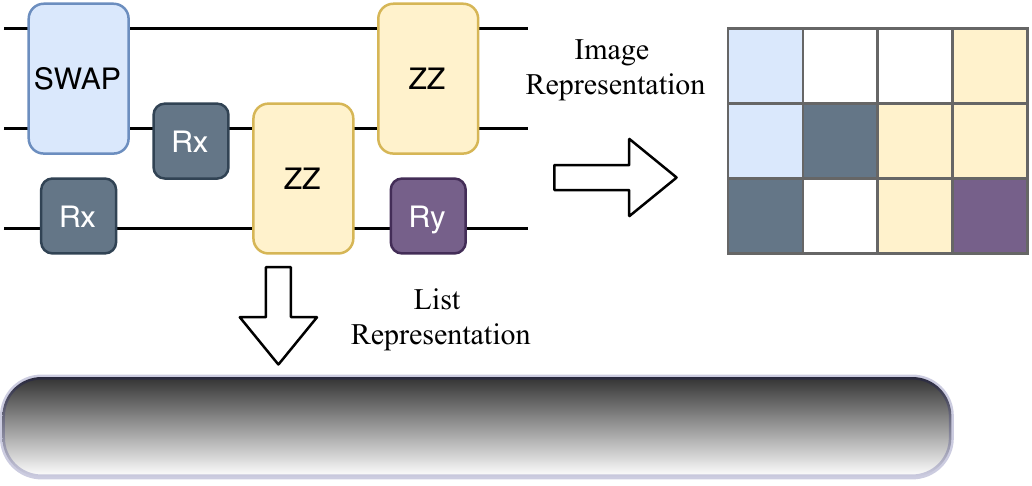}
	\caption{List representation and image representation for circuit ansatz, $N=3$, $D=4$, $n_t=6$.}
	\label{fig:imagerepr}
\end{figure}

Apart from the list representation of quantum circuits, we have alluded to the image representation which is designed for training the neural predictor. To encode the circuit structure as input for a ML predictor model, we need a systematic way to represent circuit structures in the form of tensors.  The strategy we invented is to transform a circuit to an image of  multiple channels. The shape of the input tensor is $[\#\text{depth}, \#\text{qubit}, \#\text{gate types}]$.  The size of the figure is the number of qubits times the depth of the circuit, where the depth is the number of gate layers in a circuit.  For the circuit generation pipelines considered in this work, the total number of quantum gates in a circuit $n_t$ is fixed, but different candidate circuits tend to have incompatible circuit depth. Therefore, we need to set a max depth cutoff as $D$. All circuits with depth less than $D$ are zero padded up to D columns, and all circuits with depth more than $D$ are simply retracted. In other words, we only process candidate circuits with $n_t$ quantum gates and limited to a depth of $D$.
Fig.~\ref{fig:imagerepr} gives an example of both the list representation and image representation for an ($N=3, D=4$) quantum circuit.

In our setup, this image representation maintains an one-to-one mapping with the actual quantum circuit architecture.
This is only true as we implicitly impose two restrictions on the search space: 1. the two-qubit gates are restricted to act on adjacent qubits, and 2. all two-qubit gate primitives in our work are symmetric. Namely, we do not use asymmetric two-qubit gates such as CNOT, in which two qubits play different roles as the control and the target. If the first condition is relaxed, when more than one two-qubit gates of the same type are in the same  layer of the circuit,  the image representation will have ambiguity on how to divide these qubits into pairs. For example, if in one row of the image representation (the same layer of the circuit), we have the first four elements in the same channel of SWAP gate (two SWAP gates are defined on this layer on qubits 1,2,3,4), then it is impossible to resolve whether the original circuit of this image has $\text{SWAP}_{12}, \text{SWAP}_{34}$ or $\text{SWAP}_{13}, \text{SWAP}_{24}$.
On the other hand, if the second restriction is relaxed, we then have to add more channels than the number of quantum gate types $t$.
For example, CNOT gate may require two different channels to distinguish the role of each qubit.
In the proposed QAS workflow, one may further consider different architectures of ML predictors. If an RNN based model is used, then we treat the depth dimension as the time dimension while the dimension of qubits and gate types is flattened out as an input vector for each time slice of such RNN based predictor.

Further data augmentation can be applied to the image representation of quantum circuits.
In many VQA problems, the final measurement observable is independent on the order of qubits, where a permutation on the qubit order  leaves the final result unchanged. Strictly speaking, in our search space there is no qubit permutation symmetry or redundancy since two-qubit gates are only defined on the neighboring qubits and a random permutation may break such restriction. But if we assume there is still such permutation symmetry in the representation,  input permutation on the qubits is helpful to avoid overfitting as it essentially creates $N!$ times more data than the original input.

\subsection{Architecture of predictor model}

\begin{figure}[t]
	\includegraphics[width=0.4\textwidth]{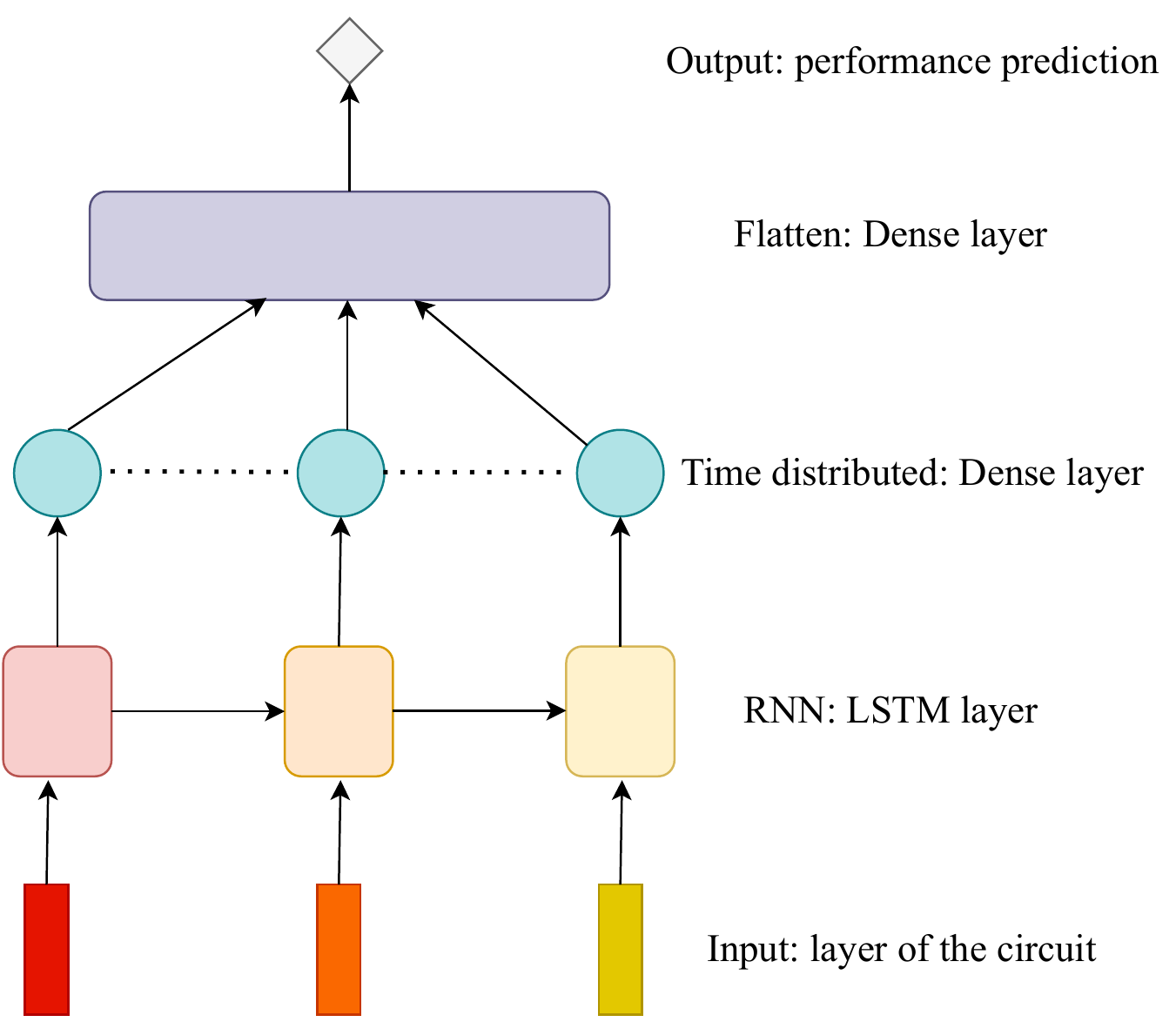}
	\caption{The schematic RNN neural architecture for the predictor of QAS. Layer of the circuit image representation are fed into the network as different time steps. The information processing flow in such network is similar to the real quantum dynamics on quantum circuit.}
	\label{fig:rnnmain}
\end{figure}

We have tried MLP, CNN and LSTM as the neural predictors to evaluate circuits in the proposed QAS workflow. In general, we find LSTM based RNN performs better than others in terms of predicting the fitness of circuit ansatz. See Fig.~\ref{fig:rnnmain} for a schematic of the RNN neural predictor used in this work.

In some VQA problems, good circuit ansatz are dense in the search space with a very long tail distribution of bad candidates.  Such a distribution is hard to fit with one regression model, and we adopt the strategy of two-stage classification for screening circuits. Firstly, a CNN based binary classification model is trained to differentiate between good and bad circuits for a task. Only good candidates are further fed into an RNN based regression model for a more fine-grained evaluation. Such regression model is only trained with good ansatz.

\subsection{Training neural predictors}

To train the neural predictor, we need to prepare a dataset composed of a circuit structure and its performance according to a task-specific evaluation metric, eg. estimated energy in VQE simulation or validation accuracy for QML. It is important that the training of such a predictor model is sample efficient, i.e. a small number of training data points should allow the predictor to deliver an acceptable accuracy. Otherwise, predictor based QAS is not desired as it already takes an enormous amount of time and resources to process the training dataset. 
In our experiments, $O(10^2)$ data pairs are in general enough to make QAS workflow a success. Of course, more training data can boost the predictor's accuracy and, in turn, elevate the overall efficiency for architecture search. There is certainly a tradeoff between the efforts to prepare the training dataset (for neural predictors) and the search efficiency in a later stage of a QAS workflow.

While preparing the training dataset, it is desired to evaluate each circuit multiple times with different initialization of parameters. As an infamous fact, the energy landscape of typical cost functions for quantum circuits is often decorated with a plethora of local minima. Therefore, one should use the minimum loss of multiple runs as the training label for the neural predictor. There are additional benefits to run the same circuit multiple times. For instance, one may train another regression model to predict the standard deviation of the losses for multiple optimization runs of a quantum circuit. This frustration indicator gives us a hint whether a candidate circuit can be consistently and easily trained for the task at hand. A prediction of large standard deviation implies the candidate circuit may suffer from a more ragged energy landscape as well as possible issue of barren plateau.  In principle, we can train a multi-task neural predictor that not only guide us for the circuit with a better potential to perform well but also easier to train from scratch.


Both mean squared loss and mean absolute loss are tried in the training of regression models, and they tend to give similar results. Adam optimizer is utilized for all the trainings in this work. Batch size is 32 or 64 in most of the training.
Dropout layers with high dropout rate are heavily applied to avoid severe overfitting.

It is worth noting that the trained neural predictors often do not perform well in terms of conventional ML evaluation metrics. The training tends to overfit even with networks of fewer parameters and large dropout ratios. However, as we will show in the Results section, such predictors are actually good enough to greatly improve the search efficiency of QAS. 

\subsection{QAS workflow}

\begin{figure}[t]
	\includegraphics[width=0.48\textwidth]{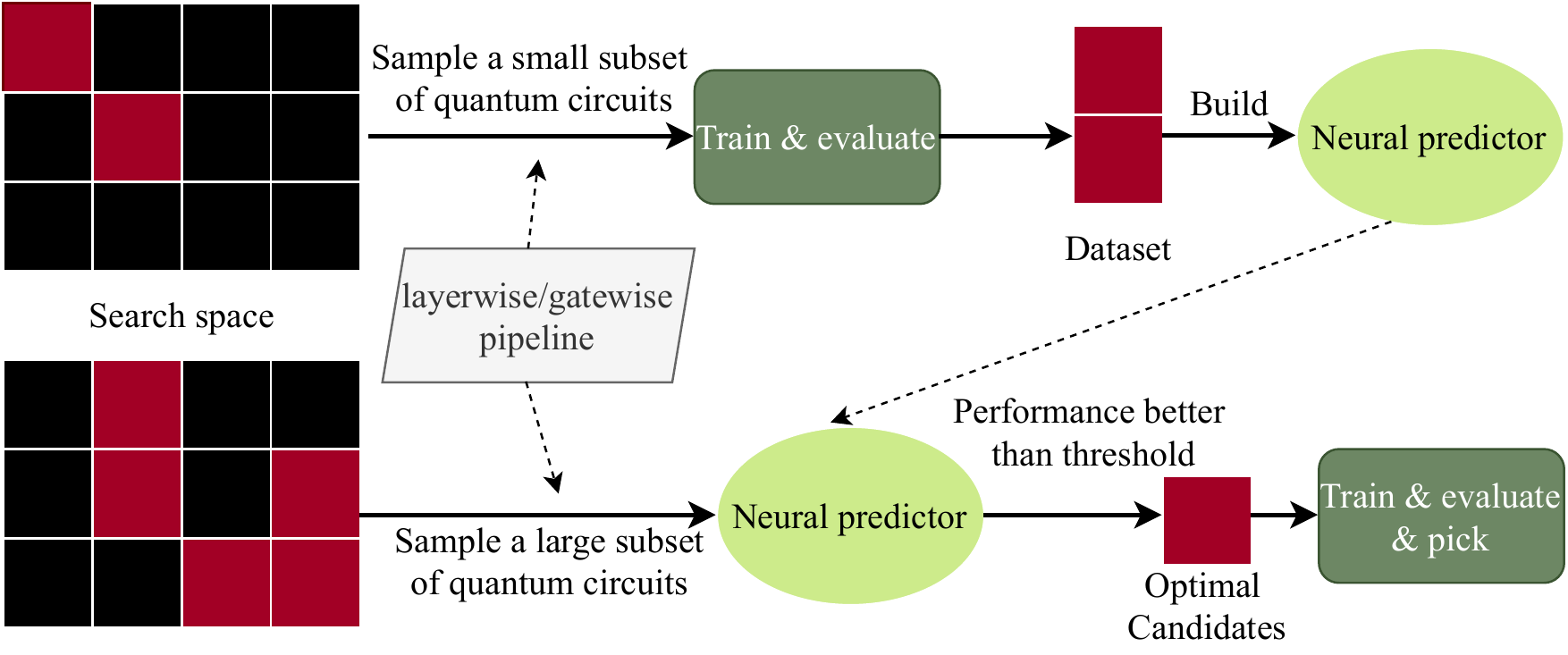}
	\caption{The workflow of predictor based QAS. There are two phases of QAS. At phase I (the upper row), we generate the dataset of quantum circuit and train a neural predictor. At phase II (the lower row), we utilize the trained predictor to filter a large number of quantum circuit candidates where only a small fraction of them with predicted performance better than a threshold is further evaluated from the scratch. The ansatz with the best performance is picked as the final result of QAS.}
	\label{fig:workflow}
\end{figure}

Once the neural predictor is trained, we randomly sample a large number of quantum circuits according to the generation pipelines we introduce.  Only circuits that pass the predictor screening (a tiny fraction of all sampled) will be actually tested to verify their fitness.  In the end, one should pick a candidate circuit of best fitness for the current task. The entire workflow of neural predictor based QAS is succinctly summarized in Fig.~\ref{fig:workflow}.


Such neural predictor based evaluation policy can be combined with more sophisticated sampling strategies to further improve search efficiency. For example, the neural predictor can be further trained with additional data collected during the Phase II of QAS workflow. One can then screen more circuits with help from a refined predictor.  This fine-tuning of neural predictor can go a few rounds iteratively and should be beneficial for finding a really strong candidate circuit. Moreover, genetic transformation or Bayesian optimization can be used as sampling strategy in the Phase II of QAS to accelerate the search. Contrary to the plain random sampling utilized in this work, these advanced sampling strategies should help as in high-throughput virtual screenings of molecules and materials. Furthermore, neural predictor combined with a VAE setup can be exploited to directly search for strong candidate circuit with gradient-based method in the continuous latent space. We leave these interesting possibilities to extend predictor based QAS workflow as promising directions to explore in a future work.

\subsection{Transferability of optimal quantum ansatz}
It is highly desirable that an optimal ansatz identified by QAS supports a transfer capacity since direct QAS on large systems is prohibitively difficult. For each optimal quantum circuit selected from QAS, we then test whether such circuit ansatz can be transferred to similar problems involving a larger number of qubits while maintaining state-of-the-art performance. Quantum circuits generated by the layerwise pipeline can be straightforwardly adapted to work on larger system by simply extending the range of each gate to cover either all odd or even qubits in a larger circuit. However, such a direct transfer of the quantum ansatz is not rooted in a rigorous theoretical analysis and may suffer from a huge performance drop.

\begin{figure}[t]\centering
	\includegraphics[width=0.48\textwidth]{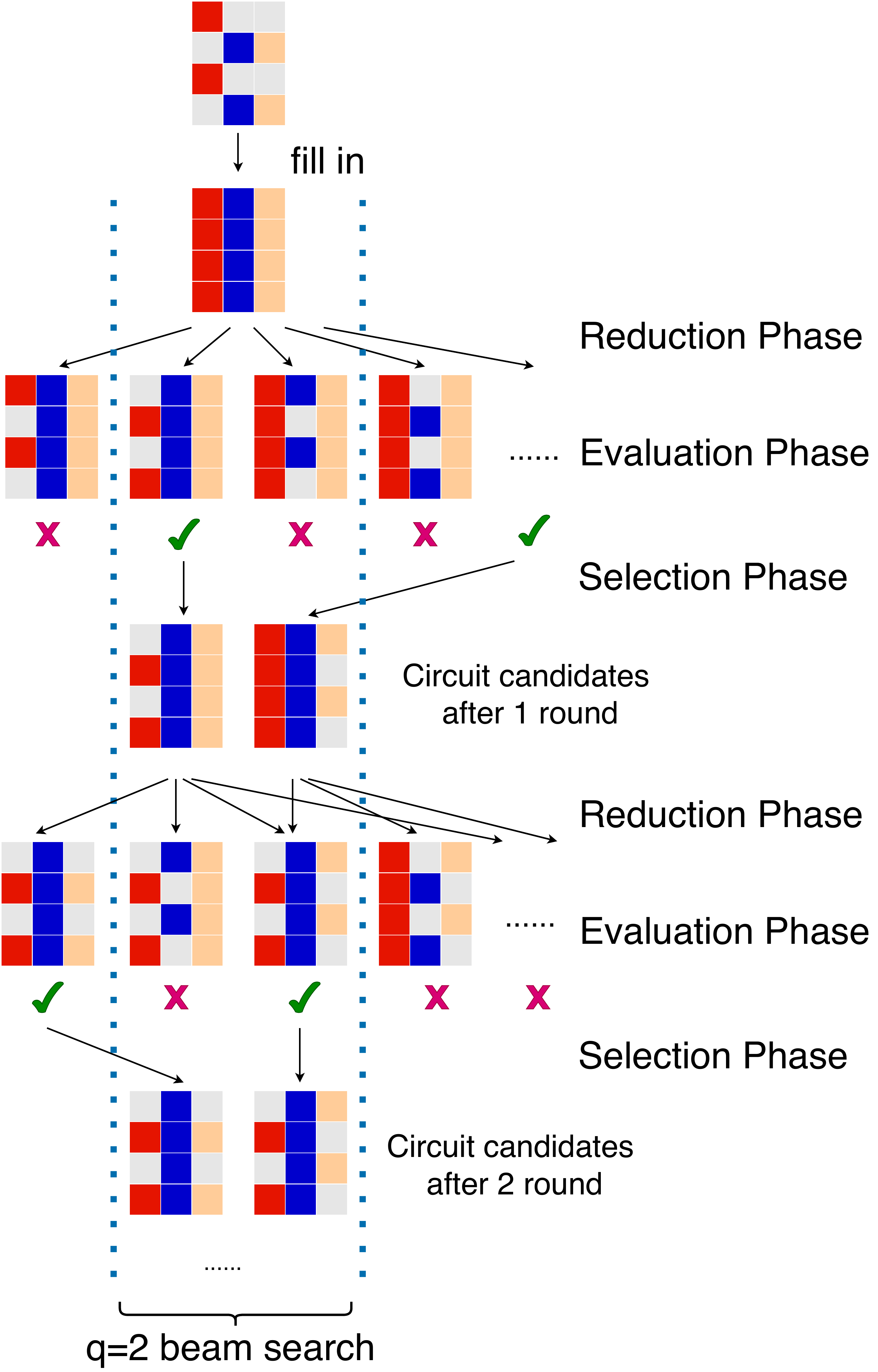}
	\caption{Schematic workflow of transferring optimal quantum ansatz. We first fill in the layerwise generated circuit structure and then do reductions by beam search ($q$=2 in this example). The circuit structure is shown as image representation, where light grey square indicates there is no quantum gate. The circuit in this example has $N=4$ qubits and depth $D=3$. Two step of the beam search is shown and more steps are possible as long as the fitness in evaluation phase is above the predefined threshold.}
	\label{fig:beamsearch}
\end{figure}

In order to find good ansatz on large size system with the knowledge gained from QAS on small systems, we develop a beam search based method to accelerate the search for appropriate ansatz for large systems while keeping the required quantum resource at a low extra overhead. Since quantum circuits generated by the layerwise pipeline always group the same quantum gate acting on either even or odd qubits in one layer, we can first fill the circuit by extending each quantum gate on all qubits (half-layer $\rightarrow$ full layer).
Such ``fill-in" quantum circuits are, in general, very good candidates with great fitness for larger size problems.
However, the new quantum circuit costs twice the quantum resource than the original circuit with half-layers. To reduce the number of gates while maintaining the performance above a threshold, we use a beam search scheme \cite{Li2020} to find simpler circuit structures with respect to a given ``fill-in" circuit. There are three phases at each step of the beam search. In the reduction phase, we reduce quantum gates of one half-layer and every possible reduction ($O(2D)$ in total) is generated. In the evaluation phase, these circuit candidates with reduction are evaluated for the fitness. Since these quantum structures can be viewed as subcircuits of the ``fill-in" one, the weights (trainable parameters) can be directly inherited and only some light fine tuning are required to get a reasonably accurate evaluation.  In the selection phase, only top-q circuits in terms of fitness is kept in the queue ($q=1$ is reduced to greedy search algorithm).  This iterative pruning of circuit structures ends when no further reduction on quantum gates is possible without compromising the expected performance. Finally, we note this technique of beam search can also incorporate mutations (or transformations) of quantum circuits in addition to eliminations of quantum gates. In this augmented version of beam search, our transfer protocol actually operates like an evolutionary algorithm for finding suitable circuits for large-size problems. See Fig~\ref{fig:beamsearch} for the workflow schematic of such fill-in + beam search protocol.

\section{Results}

In this section, we present the main results of neural predictor based QAS on two specific VQA tasks: 1. supervised learning task for binary classification on the fashion-MNIST dataset and 2. quantum simulation to estimate the ground state energy of the transverse field Ising model (TFIM).

\subsection{Quantum machine learning on classfication of fashion-MNIST}

{\bf Setup.} We train a PQC as a ML model for a binary classification task on a well-established benchmark in the ML community, and QAS is utilized to identify a suitable circuit architecture (conforming to specified constraints) that can attain rather high accuracy on the validation set. We choose fashion-MNIST \cite{Xiao2017} as the benchmark because it is more challenging than the MNIST dataset commonly tested in QML literatures. We only use the data labeled with T-shirt/top(0) and Dress(3) in order to focus on a binary classification problem instead of a multiple categorical classification. As demonstrated in \cite{Huang2020a}, QML could perform better than the classical counterparts when the training dataset is small. We, hence, select only 500 datapoints from fashion-MNIST for the training purpose, and select another 500 datapoints for validation. Each $28\times 28$ image of clothes is padding and flatten to an $1024$-dimension vector which is further encoded into a 10-qubits wavefunction using amplitude encoding.  Our quantum ansatz are hence defined in the search space with qubit number $N=10$, number of quantum primitive gates $n_t=50$ and circuit depth cutoff $D=10$. We only focus on quantum ansatz sampled from the layerwise pipeline in this QML experiment. The primitives of quantum gates include $t=7$ types in total, they are parameterized Rx, Ry, Rz, XX, YY, ZZ, and SWAP gate.

Inspired from the recent proposal of classical shadow with random measurements \cite{Huang2020b, Li2020b}, we introduce a classical post-processing module to make the actual prediction based on features extracted from random measurements on the proposed quantum circuit. 
This ``random measurement" is implemented with one extra layer of Rx gate, appended to the end of the ansatz-generation circuit.
Classical bits of information are then extracted with measurements of Pauli $Z$ operator for each qubit as $\langle Z_i \rangle$. The collected classical information is subsequently fed into a classical dense layer as $y_{pred} =\sigma(\sum_{i=1}^N k_i\langle Z_i\rangle +b)$, where $k_i$ and $b$ are classical trainable parameters and $\sigma(x) = \frac{10}{1+e^{-x}}$ is a scaled sigmoid function.
Finally, the mean squared loss between $y_{pred}$ and $y_{true}$ (the ground truth) is minimized with a simple gradient descent approach, where the weights of quantum circuit and the classical post processing module are jointly optimized. Once the training of the QML model converges, we take the classification accuracy on the validation set as the evaluation metric.


Next, we discuss details of preparing the training dataset for neural predictors. For this QML benchmark,  we adopt a loose converge condition in the phase I of QAS since the validation accuracy is known to be strongly correlated with the accuracy on the early stop. By using the early stop strategy, we may evaluate a circuit's prediction accuracy (for the fashion-MNIST data) without too many training steps.
Furthermore, we decide to train each circuit only once, because it is rather time consuming to independently re-train a QML model multiple times.  While we run the risk of getting a suboptimal estimate of a circuit's performance, we remind that our ultimate goal is to run a high-throughput screening of many quantum circuits in the phase II of a QAS workflow. Candidate circuits that pass the predictor-based screening still have to be experimentally trained and verified for their actual performances.  As long as some optimal circuit architectures are detected and passed through the screening, a QAS is deemed successful. According to this perspective, building a highly accurate neural predictor is certainly desirable but not strictly indispensable.

\begin{figure}[t]
	\includegraphics[width=0.42\textwidth]{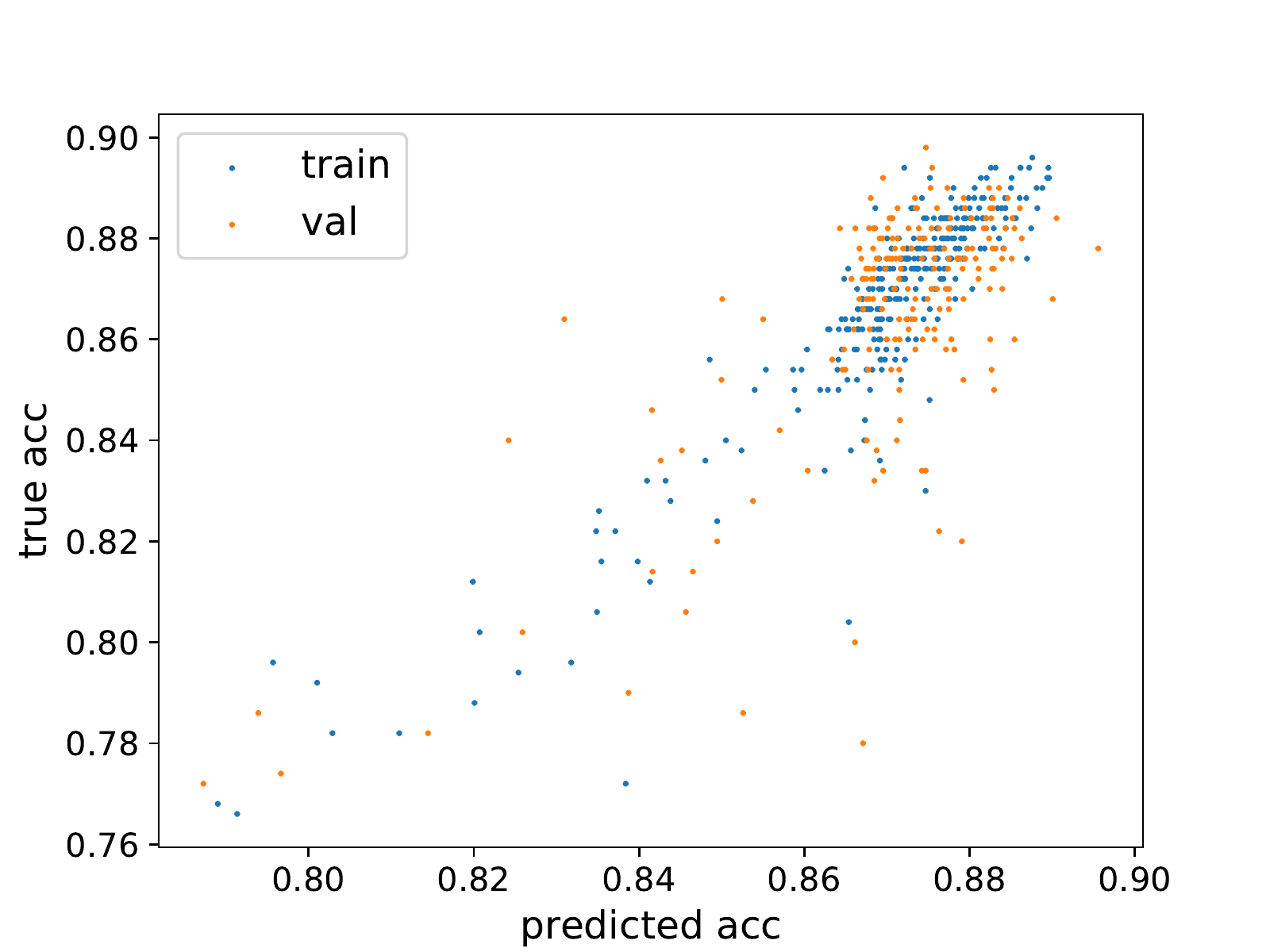}
	\caption{The regression performance of the trained neural predictor on training and validation set of quantum ansatz dataset. }
	\label{fig:qmlregression}
\end{figure}

We collect 300 datapoints for training the RNN based neural predictor. Each datapoint comprises of a quantum circuit and its corresponding  validation accuracy on the fashion-MINST as the label. The regression result is shown in Fig.~\ref{fig:qmlregression}. The performance can be evaluated by $R^2$ score of the linear regression which is around $0.71$ on the validation dataset.

\begin{figure}[t]
	\includegraphics[width=0.42\textwidth]{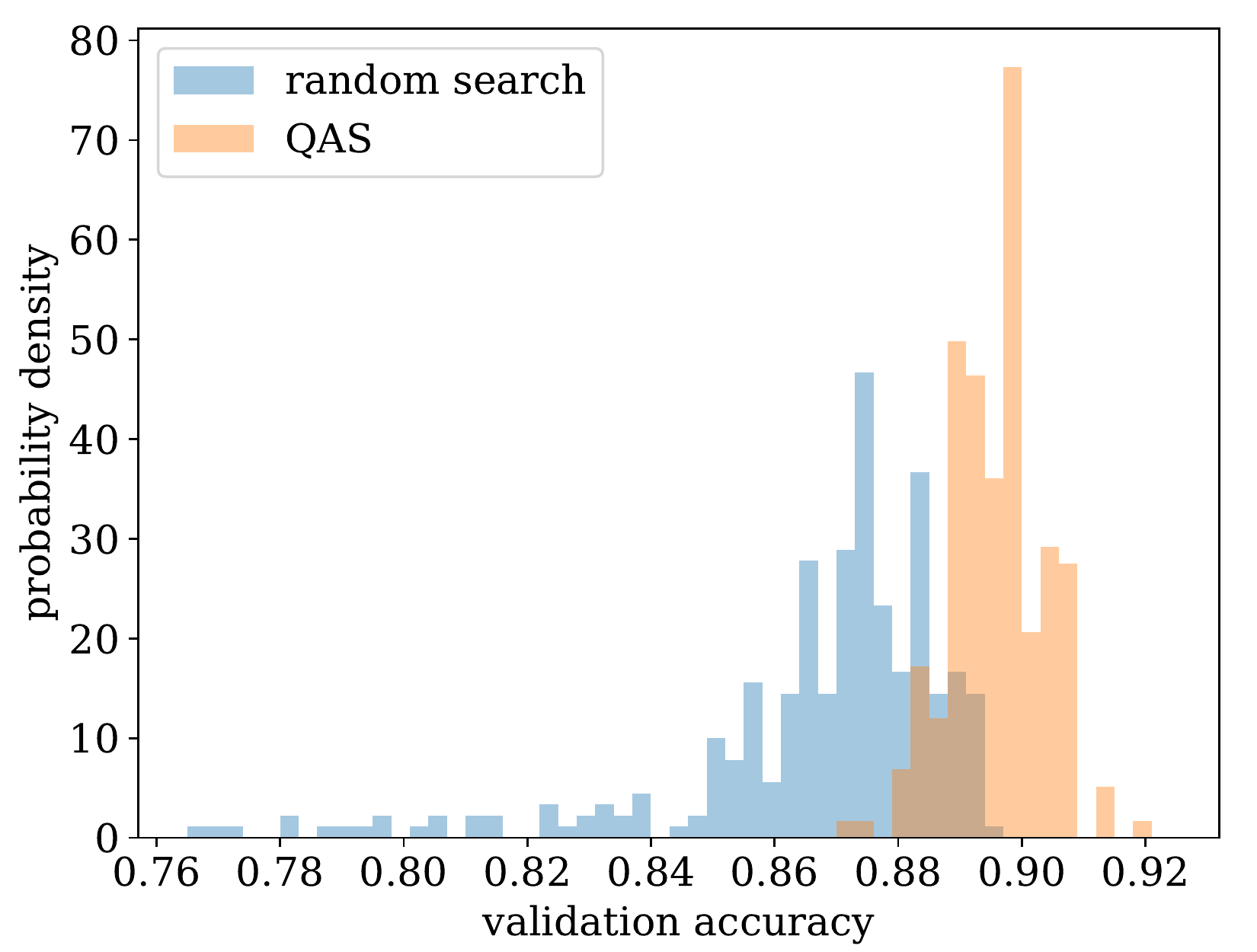}
	\caption{Validation accuracy histogram between quantum ansatz from random search and neural predictor based QAS in our work. The performance in terms of quantum machine learning classification accuracy is greatly improved in general for the set from QAS.}
	\label{fig:qashist_qml}
\end{figure}

{\bf QAS result.} The ultimate test for the QAS is how well the predictor-based screening performs in phase II. At this stage, we randomly generate quantum circuits from the layerwise pipeline, and predict their accuracy with the trained predictors. We only retain circuits with a predicted accuracy larger than $0.89$ for further training and verification of their true accuracy in experiments. For $30,000$ random circuits, only $195$ ($0.65\%$) are kept and proceeds to the actual testings.  The comparisons between the true accuracy of quantum circuits from random search (as training dataset for the predictor) and the counterpart from QAS filtered by the neural predictor is summarized in Fig.~\ref{fig:qashist_qml}. As shown, the optimal ansatz found by QAS has validation accuracy above $0.92$, which is significantly larger than the best result seen in the training set (all lower than $0.9$). The optimal quantum circuit recommended by QAS has a layerwise layout as YY, ZZ-odd, Rz-odd, YY-even, Rz-even, ZZ-even, SWAP (see Supplemental Materials for the layered ansatz notation). It is interesting to observe the generalizability of such neural predictor based approach. As the predictor is only trained with suboptimal quantum circuits with accuracy less than $0.9$, it can still single out quantum circuits with accuracy better than anyone it has seen during the training.

\begin{figure}[t]
	\includegraphics[width=0.45\textwidth]{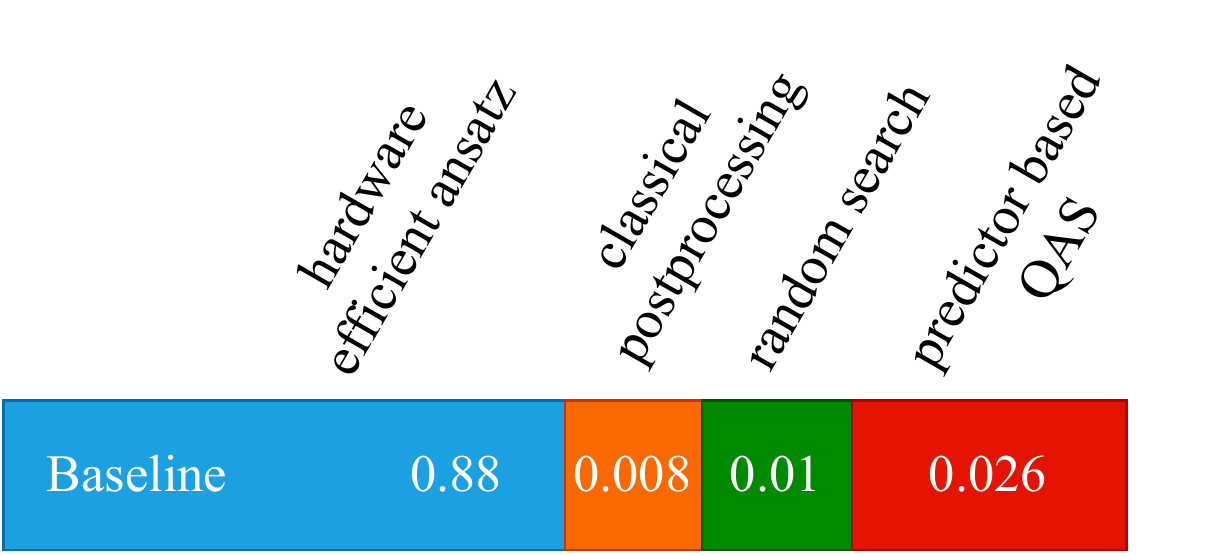}
	\caption{Validation accuracy on fashion-MNIST classification task from different contribution factors. The baseline (blue) accuracy is achieved by hardware efficient ansatz in the layer form of Rx, Ry, ZZ, Rx, Ry, ZZ.
		By attaching the classical post processing part from random measurements, the accuracy of hardware efficient ansatz get improved to $0.888$ (orange). With random search on 500 quantum circuit candidates, the best of them gives accuracy $0.898$ (green). Further QAS via the neural predictor trained from random search data record the best accuracy of $0.924$ for training on only $O(100)$ quantum circuit candidates (red).}
	\label{fig:accuracybar}
\end{figure}

We further compare QAS-screened QML model against a baseline established by a QML built with the conventional hardware efficient ansatz.
The details are given in Fig.~\ref{fig:accuracybar}.  As seen, both classical post processing and QAS contribute to the improved accuracy.

\subsection{Variational quantum eigensolver for transverse field Ising model}

{\bf Model.} Next, we investigate how predictor based QAS may help to identify an efficient state-generation circuit for representing the ground state of a many-body Hamiltonian in a VQE simulation. We consider an $N=6$ TFIM model with periodic boundary condition (PBC) for illustration. The Hamiltonian is given by $H=\sum_{i}Z_iZ_{i+1}+X_i$, and this system can be exactly modeled with 6 qubits, whose exact ground state energy is $E_0=-7.7274066$. The evaluation metric in this case is simply the best energy estimation from VQE optimization.


{\bf QAOA baseline.} We first give the baseline, a QAOA-inspired circuit, for this VQE simulation. 
This baseline circuit generalizes the QAOA structure by allowing parameters to be independently tunable intralayer. More precisely, the ansatz wavefunction reads $\prod_p\prod_i(e^{i\phi_{pi}X_i}e^{i\theta_{pi}Z_iZ_{i+1}})\prod_iH_i\vert 0\rangle$, where $p$ is the number of layers for the ansatz, all $\phi, \theta$ are trainable parameters and $H_i$ is Hadamard gate on qubit $i$. 
In the following analysis, we compare results from QAS-designed circuit against such baselines derived from this QAOA-inspired circuit with $p=1,2,3$, respectively. For the record,  these three baselines estimate the TFIM ground-state energy to be $-7.24264$, $-7.4641$, $-7.7274066$, respectively. In comparison to the true energy, $p=3$ ansatz with $42$ quantum gates and $36$ parameters can fully represent the ground state of the $N=6$ TFIM system.

{\bf Setup.} Since obtaining optimal weights for a VQE circuit usually takes less time than that for a QML task, we train each quantum circuit for $10$ independent runs with different random initializations to avoid local minimum in the estimation of the ground-state energy. These independent training can be cast into batch dimension with the help of quantum machine learning framework \cite{Broughton2020} by clever design which enables fast optimization over independent runs simultaneously. The search space for this VQE task is confined to a qubit number $N=6$, a total number of quantum primitive gates $n_t=36$ with depth cutoff $D=10$. Candidate circuits are sampled from both layerwise and gatewise generation pipelines with equal contribution. The primitive set of quantum gates include H, Rx, Ry, Rz, XX, YY, and ZZ gate this time.  Among these gates, the Hadamard gate $H$ is the only type without a tunable parameter. For the layerwise pipeline, we set $30\%$ of all generated circuits to begin with a layer of the Hadamard gate applied to all qubits as starting from the state $\vert +\rangle$ may help VQE to find better approximation to the ground state.

We adopt the two-stage screening for this VQE investigation. The rationale is that the energy distribution of the TFIM model is not smooth across a wide region in the search space; therefore, a single high-quality predictor is extremely difficult to train. To overcome this problem, we resort to using two predictors as described in Method section. First, we use a CNN based binary classification model to quickly rule out inappropriate circuits covering a wide variety of rather arbitrary circuit structures. The ``good" circuits with limited variety are then screened again with a RNN-based regression model to evaluate their performance more precisely . In this case, we use $\epsilon=(E-E_0)/14 \in [0,1)$, the normalized deviation of the estimated ground-state energy, as the predicted label.

\begin{figure}[t]
	\includegraphics[width=0.42\textwidth]{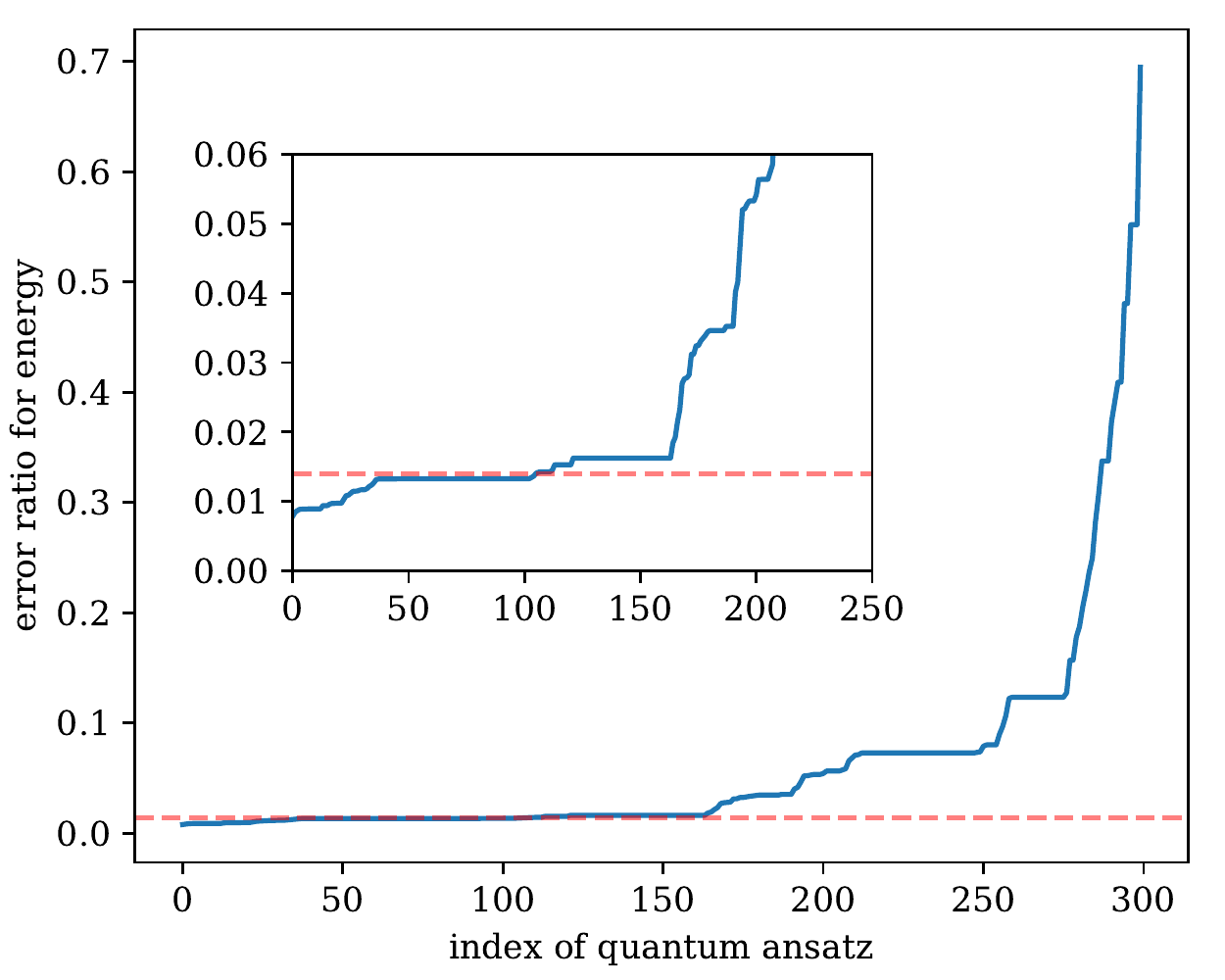}
	\caption{Sorted error ratio for VQE energy from the training dataset of quantum ansatz. The red dash line ($0.014$) is our threshold for the CNN based classification model at the first stage, i.e. we train the classification model to determine good candidates if their error ratio is less than $0.014$. The inset is the zoom-in for good quantum ansatz candidates part.}
	\label{fig:tfimdata}
\end{figure}

We again randomly pick and optimize 300 independent quantum circuits to build the training datasets for the two neural predictors.
Fig.~\ref{fig:tfimdata} shows the distribution of converged energy (error ratio $\epsilon$) for the dataset of 300 circuits.
Such a distorted distribution manifests the source of difficulty to rely on a single regression model to directly pick out top-performing circuits from the entire candidate pool. Rather, in our two-stage screening setups, the regression model is only expected to provide accurate characterization of potentially good circuit candidates with limited variety.

For the CNN based classification model, we adopt data augmentation by applying random permutations on qubits since TFIM is translational invariant and the energy is agnostic to the order of qubits. In addition, multiple layer convolution with dilation, batch normalization on qubit dimension, dropout, ELU activation and $L_2$ regularization on weights are all utilized in the classification model.

The training dataset we used only contains 300 quantum circuits and their corresponding error ratio obtained from VQE optimizations.
When one is limited to such a small training set of 300 data points, it could be crucial how the circuits are selected for representing the circuit pool for a particular problem. Prior knowledge (such as insights from physics) shall prove beneficial at this stage. For instance, only circuits conforming to certain symmetry are generated etc.  However, for the present study, we want to emphasize the universality of this predictor-based QAS and simply do a random sampling of circuits according to the two sampling pipelines. With such a scarcity of training data, the neural predictors actually do not perform particularly well according to the traditional metrics for ML model evaluation. Nevertheless, we remind that a complete QAS workflow operates more like a high-throughput screening with the ultimate goal of discovering one or few top-performing circuits as opposed to making highly accurate predictions for all circuits. Despite this focus being different from the standard ML, we still have to deal with some non-trivial challenges shared by many ML tasks. For the binary classification, we face a highly imbalanced classification with a trade-off between precision and recall.  In this case, we prefer model with higher precision instead of recall as the goal is to efficiently filter a large search space. Specifically, our trained CNN based classification model has a precision around $0.7$ and a recall around $0.47$.


{\bf QAS result.} With the two-stage predictor based screening in place, we filter a large number of circuits generated via either gatewise or layerwise pipelines. The filter threshold for the two models are correspondingly set at $0.85$ and $0.005$. We only keep the most confident candidates for further VQE evaluations.  At the first stage, only quantum circuits with predicted value larger than $0.85$ instead of $0.5$ are kept as promising candidates for further evaluation. At the second stage, only candidates with predicted error ratio less than $0.005$ will be recommended for experimental verification (i.e. going through standard VQE optimizations). It is remarkable that our RNN-based regression model can give outputs with predicted error ratio less than $0.005$ when the smallest value in its trainning dataset $0.00789$ is larger than this threshold.

\begin{figure}[t]
	\includegraphics[width=0.43\textwidth]{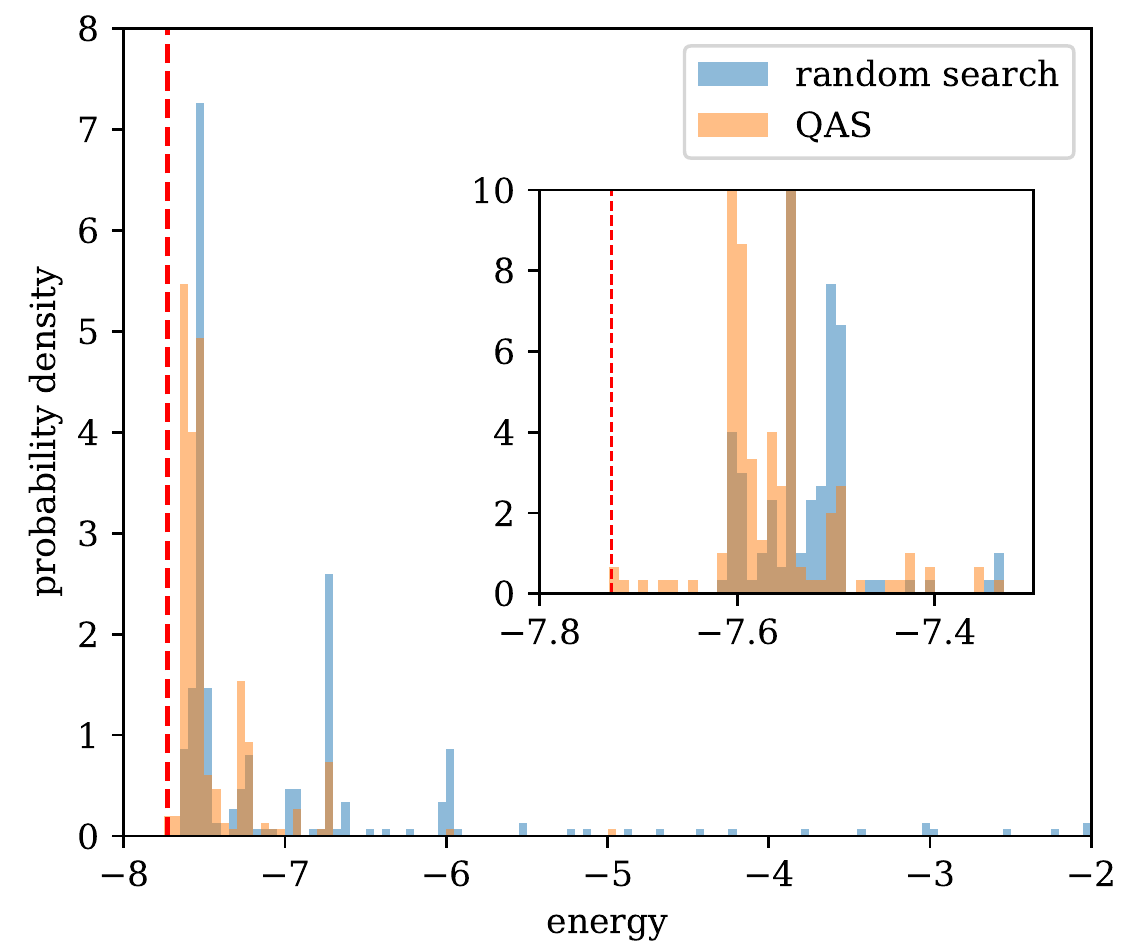}
	\caption{VQE energy histogram between quantum ansatz from random search and neural predictor based QAS. 300 samples from random search are also the training dataset of the predictor (blue). QAS can find quantum ansatz with lower energy on average and select several candidates with energy less than the best result from training datatset. The red dash line is the exact ground state energy given by exact diagonalization.
	The optimal ansatz found by QAS can indeed match the ground truth.}
	\label{fig:vqehist}
\end{figure}

We randomly sample $50,000$ quantum circuits, and only 626 ($1.25\%$) pass the two-stage screening and are actually evaluated with the VQE optimization. Among these final candidates, 5 circuits give an energy less than $-7.7$ and we denote them as optimal ansatz.
This is an interesting result as the training dataset of 300 quantum circuits do not feature such good circuits. The best energy estimation in our training dataset is only $-7.61694$. This observation demonstrates the effectiveness of our few-data trained predictors to screen unseen structures in a QAS context. In this case, the search efficiency for the predictor-based QAS workflow is around $1\%$, i.e. 1 out of every 100 quantum circuits recommended for VQE experiments are optimal (in the sense of VQE energy less than $-7.7$). Without the neural predictor as the filter, random search efficiency is around $1/2000$ following our search space pipelines according to a large number of numerical simulations. In other words, on average, predictor-based QAS only needs to conduct 100 independent VQEs before discovering an optimal candidate while naive random search requires 2000 circuits evaluation before encountering a good candidate in our pre-defined search space. This is a $20$ times boost of search efficiency. Fig.~\ref{fig:vqehist} summarizes the gain brought by the predictor based QAS compared to the random search.

{\bf Transfer of the predictor.} Next, recall that our predictors are trained with circuits limited to exactly $n_t=36$ quantum gates, but it can be transferred to predict fitness of quantum circuits consist of a distinct number of quantum gates with zero tuning. For example, we know that $p=3$ QAOA-inspired ansatz with $n_t=42$ gates can give the exact ground state. Although the total gate number is substantially changed compared to the instances in the training set  ($n_t=36$) for the predictors, the screening pipeline still works well. Classification predictor at the first stage gives the output $0.99979$ and the regression predictor gives $0.004$ as the predicted error ratio for $p=3$ QAOA-inspired circuit. In short, this optimal circuit successfully passes the two-stage screening without having to train these predictors with circuits comprising the same number of quantum gates.

\begin{figure}[t]
	\includegraphics[width=0.48\textwidth]{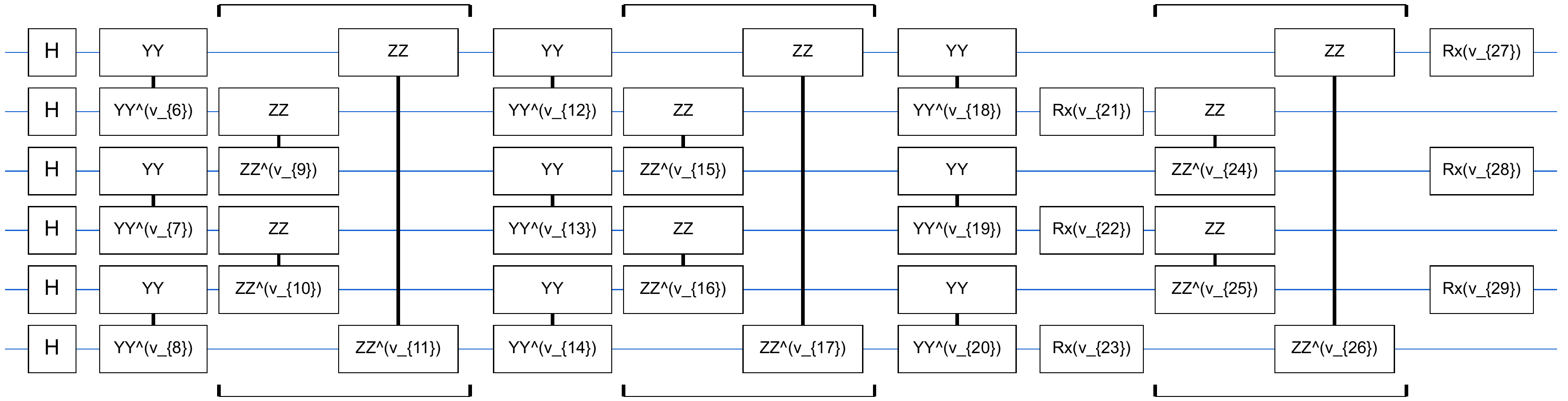}
	\caption{The optimal quantum ansatz found by neural predictor on $n_t=30$ search space. v\_\{i\} are trainable parameters for such VQE ansatz. The layerwise ansatz can be summarized as H, YY-odd, ZZ-even, YY-odd, ZZ-even, YY-odd, Rx-even, ZZ-even, Rx-odd following our convention on layerwise ansatz.}
	\label{fig:opt30}
\end{figure}

The predictors are not only able to make reasonably accurate predictions for circuits having more quantum gates but also circuits with fewer gates too. For instance, we conduct a large scale search for circuits with $n_t=30$ gates. $360$ out of $100,000$ quantum circuit structures are screened, and 3 instances of them give VQE energy smaller than $-7.7$ with the best one being  $-7.7274$.
This is a highly nontrivial result, as $p=2$ QAOA ansatz with the same amount of quantum resources only gives energy $-7.46$. The optimal circuit structure we find by transferred predictor is displayed in Fig.~\ref{fig:opt30}.

{\bf Transferability of the optimal anstaz.} Apart from the neural predictors, the QAS-designed quantum ansatz can also be transferred to guide the search for optimal circuits for similar but larger-sized problems involving more qubits in the laywerwise search space.
Such transferability is desired as it is time consuming to directly conduct QAS in a larger search space. Therefore, we propose that QAS on large systems should proceed in two steps. First, we can run QAS on smaller systems as a proxy task then we can transfer the optimal ansatz by QAS to larger systems. We conduct such transfer experiments from the optimal ansatz in Fig.~\ref{fig:opt30} to a larger TFIM model with $N=10$ spins. The exact energy and $p=3$ QAOA baseline for $N=10$ TFIM are $-12.7849$ and $-12.56758$, respectively.

We adopt the beam search approach developed in the Method section. By starting from a fully fill-in quantum circuit, we utilize the beam search to reduce potentially redundant quantum layers without compromising the final performance too much.  By this approach we obtain the following circuit structure: H-even, YY, ZZ, YY, ZZ-even, YY-odd, Rx-even. Such ansatz gives an VQE energy estimation of $-12.634$, significantly better than $p=3$ QAOA ($45$ trainable parameters in the transferred ansatz versus $60$ trainable parameters in the QAOA-inspired circuit).
If we further relax the optimal threshold, we obtain a circuit strcuture: H-even, YY, ZZ, YY Rx-even. This extremely compact anstaz containing $35$ trainable parameters and $40$ quantum primitive units outperforms the more complex $p=3$ QAOA-inspired ansatz with an VQE energy of $-12.587$. These numerical results strongly support our strategy to transfer optimal circuit structures to similar many-body simulation problems involving different number of qubits.

~\newline
\section{Discussions}

There are various future directions to refine the proposed QAS workflow and to explore other novel strategies to discover optimal quantum circuits.
One obvious possibility is to combine advanced sampling engine than the simple random search with our predictor based evaluation policy in QAS. For example, we may invoke evolutionary algorithm based sampling policy together with the learned predictor, which might further improve the search efficiency as witnessed in many high-throughput virtual screening studies.  Next, we may extend the phase II of the current QAS workflow into a loop with multiple rounds of screening. Hence, the neural predictors can be iteratively updated and fine tuned as bathes of new data points become available within each round of QAS verification of proposed circuits. Moreover, weights sharing mechanism can be combined with the predictor approach to further speed up the preparation of the training dataset.
Also, it is of great importance to further investigate transferability and propose more systematic transfer protocols for the optimal ansatz, since challenging problems are always large sized. Finally, other type of neural predictors might be helpful, such as fast indicators for quantum noise resilience \cite{Zlokapa2020} or frustration in training energy landscape. These additional considerations may be particularly crucial to establishing non-trivial quantum advantages with VQA-based approaches in the NISQ era.

{\bf Conclusion.} In this work, we introduced neural predictor as the evaluation policy for quantum architecture search.
We demonstrate the effectiveness of predictor based QAS on various examples from VQE and QML.
We find greatly improved search efficiency and new state-of-the-art quantum architectures for these VQA tasks.
Besides, we show how the trained predictor as well as the QAS-designed optimal ansatz are capable of being transferred to a different ansatz search space or problems of different size, respectively.

~\newline
{\bf Acknowledgements.} SXZ would like to thank Zhou-Quan Wan for helpful discussions. SXZ and HY are supported in part by the NSFC under Grant No. 11825404. HY is also supported in part by the MOSTC under Grant Nos. 2016YFA0301001 and 2018YFA0305604, and the Strategic Priority Research Program of Chinese Academy of Sciences under Grant No. XDB28000000.

\input{pqas_draft.bbl}%

\newpage

\begin{widetext}
	\newpage
	\section*{Supplemental Materials}
	\renewcommand{\theequation}{S\arabic{equation}}
	\setcounter{equation}{0}
	\renewcommand{\thefigure}{S\arabic{figure}}
	\setcounter{figure}{0}
	\setcounter{subsection}{0}
	\subsection{More details on search space design}
	We describe the workflow for the two pipelines for sampling circuits mentioned in the main text.
	
	The first pipeline is dubbed the gatewise generation. Given a total number of $n_t$ tuples to specify a list representation of a circuit, we have to assign the gate types and associated positions for each tuple.  To steer away from pure randomness and to keep circuits compatible with the limited connectivity of a NISQ device, the following policy is enforced. First, only two-qubit gates acting on neighboring qubits are allowed. For this study, we consider a piece of quantum hardware with a ring topology.  Next, we draw a probability logit vector from a Gaussian distribution (mean $0$, standard deviation $1.35$) for the types of gate and we increase part of the logit vector corresponding to single qubit gate by $0.5$. The motivation is to slightly favor assignments of single qubit gates since they are less susceptible to noise and easier to implement comparing to two-qubit primitives. Instead of uniformly random picking quantum gates, this hierarchical generation (first generating a probability logit vector, then generating quantum gate type according to the probability logit) restricts the variety of quantum gates appearing in a single circuit ansatz. In addition, we statistically correlate subsequent tuple choices such that, on average, more structured circuits are generated. We not only impose a higher probability to choose the same type of gates as the last one but also more likely to assign the next quantum gate in close proximity to the current one in the circuit. In our numerical investigations, circuits generated in this approach are easier to optimize and give better results overall.
	
	The second pipeline is the layerwise generation. Since, most of the times, an optimal quantum circuit possesses some higher form of regularity such as layered ansatz for problems of interests, we also sample circuits in a layer-wise fashion. Similar to the gatewise generation, we again prepare a gate-type probability logit vector. This time, whenever we pick a type of quantum gate we have to apply it on the set of even qubits or odd qubits. Namely, at each step of circuit generation, a set of tuples corresponding to a half layer of quantum gates is added to the list. Such layerwise-generated quantum circuit can also be expressed as a series of gate type. For example, ZZ, Rz-odd, YY-even represents a quantum circuit structure with ZZ gate on all sites, following with Rz gate on odd sites and YY gate on even sites (as for two-qubit quantum gates, even site means the gates are arranged on qubit (2,3), (4,5)... while odd site means the gates are arranged as (1,2), (3,4)...). Fig.~\ref{fig:pipelines} outlines the workflow of circuit ansatz generation pipelines. Finally, we emphasize that  layerwise generation pipeline is highly flexible.  One may manipulate the probability to bias toward certain circuit architecture if one has prior knowledge on certain constraints that a solution must obey. For instance, this is often the case with quantum simulation for some well-established physical models.

	\begin{figure}
		\includegraphics[width=0.66\textwidth]{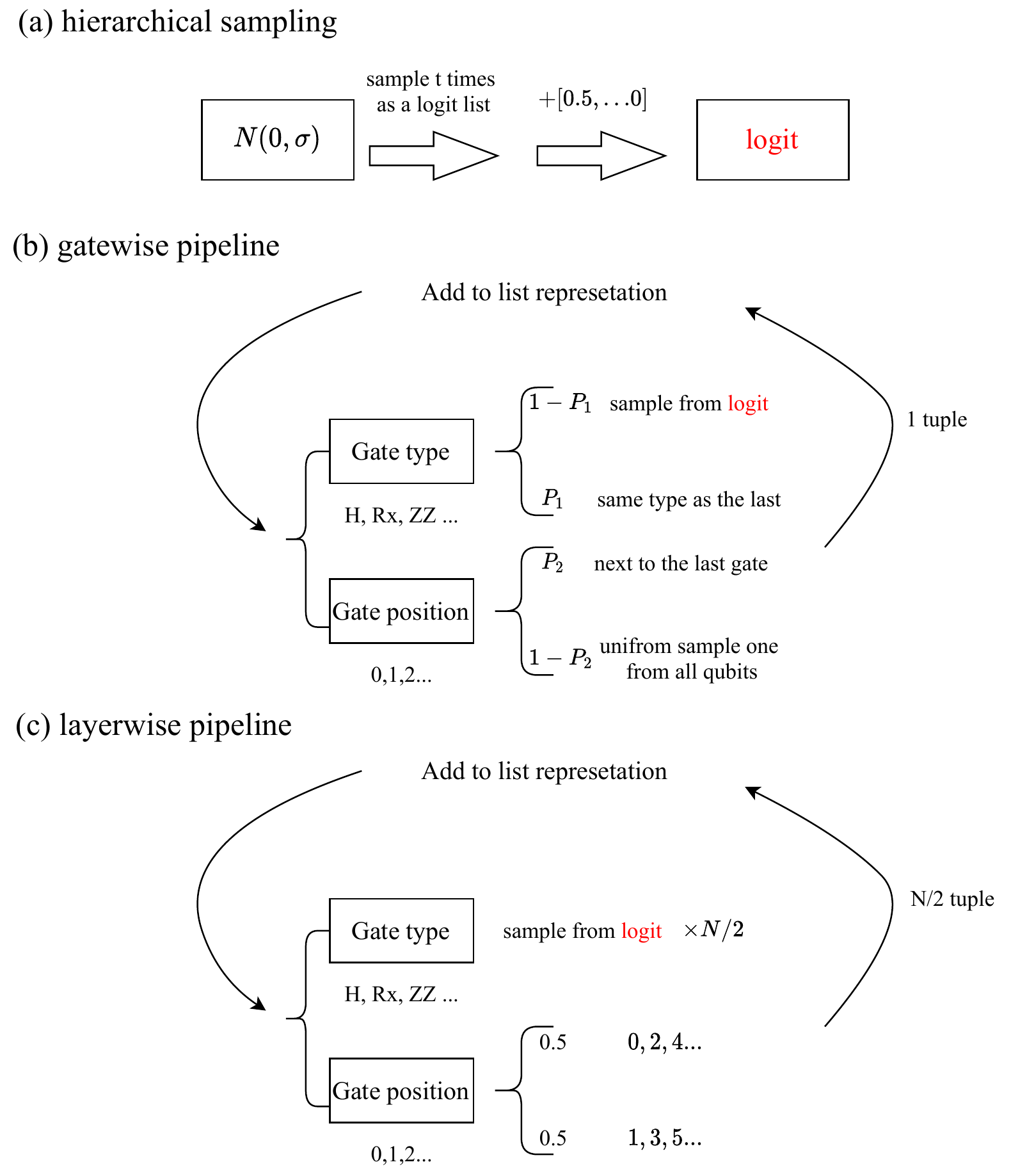}
		\caption{Search space and circuit ansatz generation pipelines in this work. (b), (c) shows the pipeline to generate new candidate ansatz in gatewise and layerwise fashion, respectively. (a) is the process to obtain probability logit for different types of quantum gate and the logit vector is used in (b), (c) when sampling quantum gate type.}
		\label{fig:pipelines}
	\end{figure}

	We emphasize the importance of designing suitable sampling pipelines. Since we sample quantum circuit candidates from such pipelines, the pipelines themselves define the legal search space as well as the distribution of circuits in this search space. For some VQA problems, such as VQE example we considered in the main text, the fitness distribution in the search space can be highly nonuniform. There is a large probability to sample a "good circuit" (say $\varepsilon<0.014$ in TFIM VQE example), but the probability to encounter a "better circuit" ($\varepsilon<0.0035$) is very rare. And as we have discussed in the main text, such probability is around $1/2000$ in the search space in VQE example. Since predictor based QAS can in general improve the search efficiency and increase such probability by only one or two order of magnitude, it is important to guarantee the probability for optimal circuit is large enough in our designed search space at the beginning. If such probability is too small, say $10^{-8}$, then the predictor based QAS still fails finding optimal circuits, as it still requires evaluation on $10^6\sim 10^7$ circuits, which is impractical. Therefore, besides QAS, the design of the search space is also of great importance, as we have to make sure optimal solution distribution is reasonably dense in the search space so that search efficiency after QAS acceleration is acceptable.

	\subsection{Neural architecture for predictor models}
	See Fig.~\ref{fig:model} for the neural network models we utilized as predictors in QAS.
	
	\begin{figure}[h]
		\subfloat[]{\label{fig:cnn}
			\includegraphics[width=0.3\textwidth]{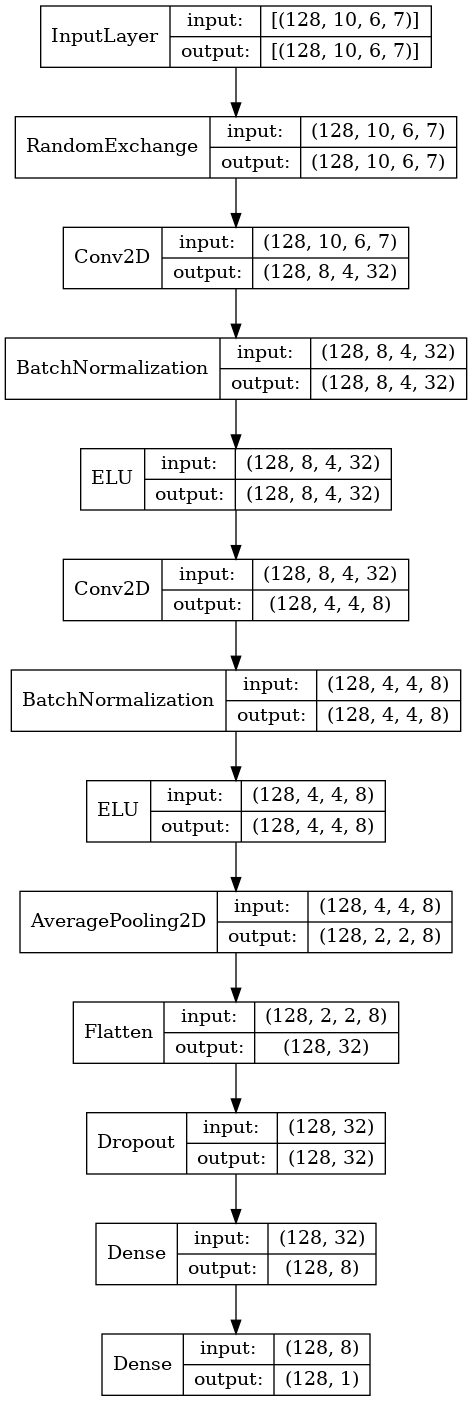}}~~~~~
		\subfloat[]{\label{fig:rnn}
			\includegraphics[width=0.39\textwidth]{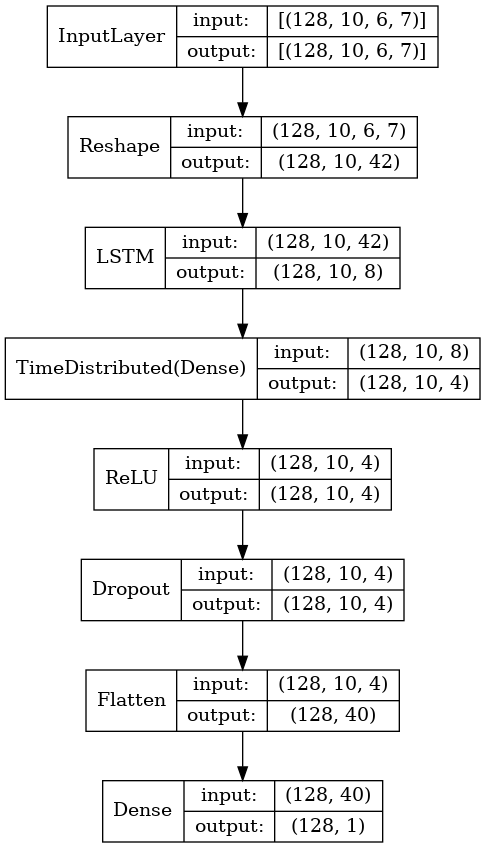}}
		
		\caption{(a) CNN based model for classification at the first stage of the predictor pipeline. (b) RNN(LSTM) based model for regression at the second stage of the predictor pipeline. $[10,6,7]$ is the shape of $N=6$, $D=10$ and gate type $t=7$, consistent with VQE task we investigated. RandomExchange is the data augmentation layer which permutes the qubit order.
		}\label{fig:model}
	\end{figure}
\end{widetext}
	
\end{document}

%% file: pqas_draft.bbl
%